\documentstyle[12pt]{article}
 
\begin{document}

\begin{titlepage}

\begin{flushright}
IJS-TP-97/11\\
\end{flushright}

\vspace{.5cm}

\begin{center}
{\Large \bf Semileptonic and nonleptonic charmed meson decays 
in an effective model}
\footnote{Contributed paper for LEPTON - PHOTON INTERACTIONS, Hamburg, 
Germany, 28 July - 1 August 1997}\\

\vspace{1.5cm}

{\large B. Bajc $^{a,b}$, S. Fajfer $^{a}$, 
R. J. Oakes $^{c}$ and S. Prelov\v sek $^{a}$}

\vspace{.5cm}

{\it a) J. Stefan Institute, Jamova 39, P.O.Box 3000, 
1001 Ljubljana, Slovenia}

\vspace{.5cm}

{\it b) Department of Physics, University of Durham, Durham, DH1 3LE, Great 
Britain}

\vspace{.5cm}

{\it c) Department of Physics and Astronomy
Northwestern University, Evanston, Il 60208
U.S.A.}

\vspace{1cm}

\end{center} 

\centerline{\large \bf ABSTRACT}

\vspace{0.5cm}

We analyze charm meson semileptonic $D \to V  l 
\nu_l$ and $D\to P l \nu_l$ and nonleptonic 
$D \to P V$, $D \to PP$ and $D \to VV$ decays 
within a model which combines 
the heavy quark effective Lagrangian and chiral perturbation theory.  
\end{titlepage}

\centerline{\bf I. INTRODUCTION}
 
\vskip 1cm

The experimental data for the semileptonic decays of 
D mesons are unfortunately 
not good enough to clearly determine the $q^2$ dependence 
of the form factors. What is known experimentally, apart from 
the branching ratios, are the form factors at one kinematical point, 
assuming a pole - type behavior for all the form factors. 
This assumption seems reasonable, but within heavy quark effective 
theory (HQET) the kinematic constrait on the 
form factors at $q^2 = 0$ cannot be satisfied unless a 
special relation is imposed between the pole masses and residues. 
Recently we have developed  a model  for the semileptonic decays 
$D \to V l \nu_l $  and $D \to P l \nu_l $, where $P$ and $V$ are light 
$J^P=0^-$ and $1^-$ mesons,  respectively \cite{BFO0}. 
This model combines the heavy quark effective theory  
and the chiral Lagrangians. 
HQET is valid at a small recoil momentum \cite{CAS1,WISE}   
and can give definite
predictions for heavy to light ($D\to V$ or $D\to P$)
semileptonic decays in the kinematic region
with large momentum transfer $q^2$ to the lepton pair.  
Unfortunately, it cannot predict the $q^2$ dependence of the 
form factors \cite{CAS1,WISE}.
For these reasons, we have modified the Lagrangian for
heavy and light pseudoscalar and vector mesons
given by the HQET and chiral symmetry \cite{CAS1}.
Our model \cite{BFO0} gives a natural explanation of the
pole-type form factors in the whole $q^2$ range,  
and it determines which form factors have a
pole - type or a constant behaviour, confirming the
results of the QCD sum rules analysis \cite{BALL}.
To demonstrate that this model  works well,
we have calculated the decay widths in all measured charm
meson semileptonic decays \cite{BFO0}.  The model parameters were 
determined by the experimental values  of two measured 
semileptonic decay widths. 
The predictions of the model are in good agreement 
with the remaining experimental data on semileptonic decays.
 
The nonleptonic D meson decays are challenging to understand 
theoretically  (see e.g. \cite{BFOP} and references therein). 
The short distance effects are now well understood \cite{BURAS},
but the nonperturbative techniques required for 
the evaluation of certain matrix elements are based on approximate 
models. Often the factorization approximation is used 
(see e.g.\cite{BFOP} and references therein). 
The amplitude for the nonleptonic weak decay is then considered 
as a sum of the ``spectator'' contribution 
and the ``annihilation'' contribution,  
the direct annihilation of the initial heavy meson.
In the determination of the ``spectator'' contribution one uses the 
knowledge of the hadronic matrix elements calculated in 
D meson semileptonic decays. 
Another problem in the analysis of nonleptonic D meson decays is  the 
final state interactions (FSI) \cite{BLMP,BLMMPS,BLMPS,BLP,WSB,WSB1}. 
These arise from the interference of different isospin states 
or the presence of intermediate resonances, and both 
spectator and annihilation amplitudes can be affected. 
The FSI are especially important for the annihilation contribution, 
which can often 
be  successefully described by the dominance of nearby scalar 
or pseudoscalar resonances \cite{BLMP,BLMMPS,BLMPS,BLP}. 
The effective model developed to describe the 
$D \to V(P) l \nu_l$ decay widths \cite{BFO0}
contains only light vector and pseudoscalar final states and, therefore,  
is not applicable to the annihilation amplitudes.   
Consequently, in the present paper we apply  this  effective model 
to analyze only those $D\to PV$, $D\to PP$, and $D\to VV$ decays in 
which the annihilation amplitude is absent or negligible. 
Other FSI might arise as a result of elastic or inelastic rescattering. 
In this case, the two body nonleptonic D meson decay amplitudes can 
be written in terms of isospin amplitudes and strong interaction phases 
\cite{KP}. 
As usual, we assume that the  important contributions to FSI are 
included in 
these phases. In fact, we will avoid the effects of the FSI strong 
interaction phases by considering only the D meson decay modes in which 
the final state involves only a 
single isospin. Our analysis then includes  the decays 
$D^+ \to \bar 
K^{*0} \pi^+$, $D^+ \to \rho^+ \bar K^{0}$, $D^+ \to \bar K^0 \pi^+$, 
$D^+ \to \bar K^{*0} \rho^+$, $D^+ \to \Phi \pi^+$, 
$D^+_s \to \Phi \pi^+$, $D_s^+ \to \Phi \rho^+$, $D^0 \to \Phi 
\omega^0$, $D^0 \to \Phi \eta$, $D^+ \to \rho^+ \eta (\eta ')$ 
and $D^0 \to \omega^0 \eta (\eta ')$.

To evaluate the  spectator graphs for nonleptonic decays  we use 
the form factors for the $D\to V$ and $D\to P$ weak decays, 
calculated for the semileptonic decays \cite{BFO0}.  This explores 
how well their  particular $q^2$ behavior also explains 
the nonleptonic decay amplitudes. 
At the same time the analysis of 
the nonleptonic decays enables us 
to choose between different solutions for the model 
parameters found in the semileptonic decays, determining the set of the 
solutions which are in the best agreement with the experimental results 
for the nonleptonic decay widths. Moreover, we obtain a value 
for the parameter $\beta$, which 
can not be determined from the semileptonic decay alone, 
but enters in the nonleptonic decays.

The paper is organized as follows. In Sec. II we present the effective 
Lagrangian for heavy and light pseudoscalar and vector mesons, 
determined by the requirements of HQET and chiral symmetry. 
In Sec. III we present the results for the 
$D \to V  l \nu_l$, $D \to P l \nu_l$  decays \cite{BFO0}. 
In Sec. IV we analyze the nonleptonic decay widths. 
Finally, a short summary of the results is given in Sec. V.

\vskip 1cm

\centerline{\bf II. THE HQET AND CHPT LAGRANGIAN FOR 
$D \to V (P) l \nu$ }

\vskip 1cm

We incorporate in our Lagrangian
both the heavy flavour $SU(2)$ symmetry, 
and the $SU(3)_L\times SU(3)_R$ chiral
symmetry, spontaneously broken to the diagonal
$SU(3)_V$ \cite{BFO0} (and references therein), which can be used for the
description of heavy and light pseudoscalar and
vector mesons. 
The light degrees of freedom are described by the
3$\times$3 Hermitian matrices

\begin{eqnarray}
\label{defpi}
\Pi = \pmatrix{
{\pi^0\over\sqrt{2}}+{\eta_8\over\sqrt{6}}+{\eta_0\over\sqrt{3}} &
\pi^+ & K^+ \cr
\pi^- & {-\pi^0\over\sqrt{2}}+{\eta_8\over\sqrt{6}}+
{\eta_0\over\sqrt{3}} & K^0 \cr
K^- & {\bar K^0} & -{2 \over \sqrt{6}}\eta_8+
{\eta_0\over\sqrt{3}} \cr}\;,
\end{eqnarray}

\noindent
and

\begin{eqnarray}
\label{defrho}
\rho_\mu = \pmatrix{
{\rho^0_\mu + \omega_\mu \over \sqrt{2}} & \rho^+_\mu & K^{*+}_\mu \cr
\rho^-_\mu & {-\rho^0_\mu + \omega_\mu \over \sqrt{2}} & K^{*0}_\mu \cr
K^{*-}_\mu & {\bar K^{*0}}_\mu & \Phi_\mu \cr}
\end{eqnarray}

\noindent
for the pseudoscalar and vector mesons, respectively.
The mass eigenstates are defined by
$\eta=\eta_8\cos{\theta_P}-\eta_0\sin{\theta_P}$ and
$\eta'=\eta_8\sin{\theta_P}+\eta_0\cos{\theta_P}$, where
$\theta_P=(-20\pm 5)^o$ \cite{PDG} is the $\eta-\eta'$
mixing angle.
The matrices (\ref{defpi}) and (\ref{defrho}) are conveniently written in 
terms of 

\begin{eqnarray}
\label{defu}
u & = & \exp  ( \frac{i \Pi}{f} )\;,
\end{eqnarray}

\noindent
where $f$ is the pseudoscalar decay constant, and

\begin{eqnarray}
\label{defrhohat}
{\hat \rho}_\mu & = & i {g_V \over \sqrt{2}} \rho_\mu\;,
\end{eqnarray}

\noindent
where $g_V=5.9$ is given by the values of the
vector masses since we assume the exact
vector dominance \cite{BFO0}.
Introducing the vector and axial currents
${\cal V}_{\mu} =  \frac{1}{2} (u^{\dag}
\partial_{\mu} u + u \partial_{\mu}u^{\dag})$
and ${\cal A}_{\mu}  =  \frac{1}{2} (u^{\dag}
\partial_{\mu} u - u \partial_{\mu}u^{\dag})$
and the gauge field tensor
$F_{\mu \nu} ({\hat \rho}) =
\partial_\mu {\hat \rho}_\nu -
\partial_\nu {\hat \rho}_\mu +
[{\hat \rho}_\mu,{\hat \rho}_\nu]$
the light meson part of the strong
Lagrangian can be written as

\begin{eqnarray}
\label{defllight}
{\cal L}_{light} = &-&{f^2 \over 2}
\{tr({\cal A}_\mu {\cal A}^\mu) +
2\, tr[({\cal V}_\mu - {\hat \rho}_\mu)^2]\}\nonumber\\
& + & {1 \over 2 g_V^2} tr[F_{\mu \nu}({\hat \rho})
F^{\mu \nu}({\hat \rho})]\;.
\end{eqnarray}

Both the heavy pseudoscalar and the heavy vector
mesons are incorporated in the $4\times 4$ matrix

\begin{eqnarray}
\label{defh}
H_a& = & \frac{1}{2} (1 + \!\!\not{\! v}) (D_{a\mu}^{*}
\gamma^{\mu} - D_{a} \gamma_{5})\;,
\end{eqnarray}

\noindent
where $a=1,2,3$ is the $SU(3)_V$ index of the light
flavours and $D_{a\mu}^*$ and $D_{a}$ annihilate a
spin $1$ and spin $0$ heavy meson $c \bar{q}_a$ of
velocity $v$, respectively. They have a mass dimension
$3/2$ instead of the usual $1$, so that the Lagrangian
is explicitly mass independant in the heavy quark
limit $m_c\to\infty$. Defining

\begin{eqnarray}
\label{defhbar}
{\bar H}_{a} & = & \gamma^{0} H_{a}^{\dag} \gamma^{0} =
(D_{a\mu}^{* \dag} \gamma^{\mu} + D_{a}^{\dag} \gamma_{5})
\frac{1}{2} (1 + \!\!\not{\! v})\;,
\end{eqnarray}

\noindent
we can write the leading order strong  Lagrangian as

\begin{eqnarray}
\label{deflstrong}
{\cal L}_{even} & = & {\cal L}_{light} +
i Tr (H_{a} v_{\mu} (\partial^{\mu}+{\cal V}^{\mu})
{\bar H}_{a})\nonumber\\
& + &i g Tr [H_{b} \gamma_{\mu} \gamma_{5}
({\cal A}^{\mu})_{ba} {\bar H}_{a}]
 +  i \beta Tr [H_{b} v_{\mu} ({\cal V}^{\mu}
- {\hat \rho}^{\mu})_{ba} {\bar H}_{a}]\nonumber\\
& + &  {\beta^2 \over 4 f^2 }
Tr ({\bar H}_b H_a {\bar H}_a H_b)\;.
\end{eqnarray}
This Lagrangian contains two unknown parameters,
$g$ and $\beta$, which are not determined by symmetry
arguments, and must be determined empirically.
This is the most general even-parity Lagrangian
of leading order in the heavy quark mass
($m_Q\to\infty$) and the chiral symmetry limit
($m_q\to 0$ and the minimal number of derivatives).

We will also need the odd-parity Lagrangian for the
heavy meson sector. The lowest order contribution
to this Lagrangian is given by

\begin{eqnarray}
\label{defoddheavy}
{\cal L}_{odd} & = & i {\lambda} Tr [H_{a}\sigma_{\mu \nu}
F^{\mu \nu} (\hat \rho)_{ab} {\bar H_{b}}]\;.
\end{eqnarray}

\noindent
The parameter $\lambda$ is free, but we know that
this term is of the order $1/\Lambda_\chi$ with
$\Lambda_\chi$ being the chiral perturbation theory
scale.

\vspace{1cm}

\centerline{\bf III. FORM FACTORS IN $D \to V/ P \nu_l l$ DECAYS}

\vspace{1cm}

For the semileptonic decays the weak Lagrangian is given at 
the quark level by the current - current Fermi interaction

\begin{equation}
\label{wl}
{\cal L}_{eff} (\Delta C = \Delta S = 1) = - \frac{G_F}{{\sqrt 2}}
[{\bar l}\gamma_{\mu} (1- \gamma_5) \nu_l 
{\bar s^{\prime}}\gamma^{\mu} (1- \gamma_5) c]
\end{equation}

\noindent
where $G_F$ is the Fermi constant, and 
$s^{\prime} =s cos \theta_C + d sin \theta_C$, $\theta_C$ 
being the Cabibbo angle.

At the meson level we 
assume that the weak current transforms as $({\bar 3}_L,1_R)$ under 
chiral $SU(3)_L \times SU(3)_R$  and is linear in the heavy meson fields. 
In our calculation of the $D$ meson semileptonic decays to
leading order in both $1/M$ and the chiral expansion we have 
shown  that the weak current is \cite{BFO0} 

\begin{eqnarray}
\label{j}
{J}_{a}^{\mu} = &\frac{1}{2}& i \alpha Tr [\gamma^{\mu}
(1 - \gamma_{5})H_{b}u_{ba}^{\dag}]\nonumber\\
&+& \alpha_{1}  Tr [\gamma_{5} H_{b} ({\hat \rho}^{\mu}
- {\cal V}^{\mu})_{bc} u_{ca}^{\dag}]\nonumber\\
&+&\alpha_{2} Tr[\gamma^{\mu}\gamma_{5} H_{b} v_{\alpha}
({\hat \rho}^{\alpha}-{\cal V}^{\alpha})_{bc}u_{ca}^{\dag}]+...\;,
\end{eqnarray}

\noindent
where $\alpha=f_D\sqrt{m_D}$ \cite{WISE}. The $\alpha_1$ term was
first considered in \cite{CAS1}. We found \cite{BFO0} 
that the $\alpha_2$ 
gives a contribution of the same order in $1/M$ and the chiral
expansion as the term proportional to $\alpha_1$.
 
The $H\to V$ and $H\to P$ current matrix
elements can be quite generally written as

\begin{eqnarray}
\label{defhv}
<V_{(i)}(\epsilon,p')|(V-A)^\mu|H(p)>=
-{2 V^{(i)}(q^2)\over m_H+m_{V(i)}}
\epsilon^{\mu\nu\alpha\beta}\epsilon_\nu^* p_\alpha
{p'}_\beta \nonumber\\
-i \epsilon^*.q {2 m_{V(i)}\over q^2}q_\mu A^{(i)}_{0}(q^2)
+i(m_H+m_{V(i)})(\epsilon_\mu^*-
{\epsilon^*.q\over q^2}q_\mu)A^{(i)}_{1}(q^2) \nonumber\\
-{i\epsilon^*.q\over m_H+m_{V(i)}}\biggl[ (p+p')_\mu-
{m_H^2-m_{V(i)}^2\over q^2}q_\mu\biggr] A^{(i)}_{2}(q^2)\;,
\end{eqnarray}

\noindent
and

\begin{eqnarray}
\label{defhp}
<P_{(i)}(p')|(V-A)_\mu|H(p)>&=&[(p+p')_\mu-
{m_H^2-m_{P(i)}^2\over q^2}q_\mu]F^{(i)}_{1}(q^2)\nonumber\\
&+&{m_H^2-m_{P(i)}^2\over q^2}q_\mu F^{(i)}_{0}(q^2)\;,
\end{eqnarray}

\noindent
where, $q=p-p'$ is the exchanged momentum and the index  $(i)$ specifies 
the particular final meson, $P$ or  $V$. In order
that these matrix elements be finite at $q^2=0$,
the form factors must satisfy the relations

\begin{equation}
\label{relff}
A_0(0)+{m_H+m_{V}\over 2 m_{V}}A_1(0)-
{m_H-m_{V}\over 2 m_{V}}A_2(0)=0\;.
\end{equation}

\begin{equation}
\label{f1f0}
F_1(0)=F_0(0)\;.
\end{equation}

\noindent
and, therefore, are not free parameters.

In order to extrapolate the amplitude
from the zero recoil point to the rest of the
allowed kinematical region we have made a very simple,
physically motivated, assumption: 
{\it the vertices do not change significantly, while the 
propagators of the off-shell heavy mesons are given by
the full propagators $1/(p^2-m^2)$ instead of the HQET
propagators $1/(2 m v \cdot k)$} \cite{BFO0}. With these assumptions
we are able to incorporate the following features: 
 the HQET prediction almost exactly at the maximum $q^2$;
a natural explanation for the pole-type
form factors when appropriate;
 and predictions of flat $q^2$ behaviour for the form factors
$A_1$ and $A_2$, which has been confirmed in the QCD sum-rule
analysis of \cite{BALL}.

Finally, we include $SU(3)$ symmetry breaking by
using the physical mas\-ses and decay constants
shown in Table 1 of ref. \cite{BFO0}. 
The decay constants for the $\eta$ and $\eta'$ were taken from
\cite{ETA}, for the 
light vector mesons from \cite{BLMPS} and for the $D$ mesons from 
\cite{RICH}, \cite{MARTIN} and \cite{BFO4}.

The relevant form factors for $D \to V$ decays defined in (\ref{defhv}) 
calculated
in our model \cite{BFO0}, are 

\begin{eqnarray}
\label{v}
{1\over K_{V(i)}} V^{(i)}(q^2)&=&(m_H+m_{V(i)})
\biggl(2{m_{H'^*(i)}\over m_H}\biggr)^{1\over 2}
{m_{H'^*(i)} \over q^2-m_{H'^*(i)}^2} 
f_{H'^*(i)}\lambda {g_V\over\sqrt{2}}\;\\
\label{a0}
{1\over K_{V(i)}} A^{(i)}_{0}(q^2)&=&\Big[{1\over m_{V(i)}}
\biggl({m_{H'(i)}\over m_H}\biggr)^{1\over 2}
{q^2 \over q^2-m_{H'(i)}^2}f_{H'(i)}\beta\nonumber\\
&+&{\sqrt{m_H}\over
m_{V(i)}}\alpha_1  - {1\over 2}
{q^2+m_H^2-m_{V(i)}^2\over m_H^2}
{\sqrt{m_H}\over m_{V(i)}} \alpha_2\Big]{g_V\over\sqrt{2}}\;,\\
\label{a1}
{1\over K_{V(i)}} A^{(i)}_{1}(q^2)&=&-{2\sqrt{m_H}\over m_H+m_{V(i)}}
\alpha_1{g_V\over\sqrt{2}}\;\\
\end{eqnarray}
and
\begin{eqnarray}
\label{a2}
{1\over K_{V(i)}} A^{(i)}_{2}(q^2)&=&\Big[-{m_H+m_{V(i)} 
\over m_H\sqrt{m_H}}\alpha_2\Big]{g_V\over\sqrt{2}}\;,
\end{eqnarray}

\noindent
where the pole mesons and the constants
$K_{V(i)}$, which contribute to the  corresponding processes 
$D \to PV$ and $D \to V_{(1)}V_{(2)}$  are given in \cite{BFO0}. 

We determined the three parameters ($\lambda$, $\alpha_1$,
$\alpha_2$) in \cite{BFO0} using the three measured values 
of helicity amplitudes $\Gamma/\Gamma_{TOT}=0.048\pm 0.004$,
$\Gamma_L/\Gamma_T=1.23\pm 0.13$ and
$\Gamma_+/\Gamma_-=0.16\pm0.04$ for the process
$D^+\to\bar{K}^{*0} l^+ \nu_l$,
taken from the Particle Data Group
average of all the data \cite{PDG}.
The parameter $\beta$ could not be determined 
from this decay rate, since $A_0(q^2)$ cannot be observed in the 
semileptonic 
decays. 

The model parameters appear linearly in the form factors
(\ref{v})-(\ref{a2}),
so the polarized decay rates $\Gamma_0$, $\Gamma_+$ and $\Gamma_-$ 
are quadratic functions of them. For this reason
there are $8$ sets of solutions for the three parameters
($\lambda$,$\alpha_1$,$\alpha_2$). It was found
from the analysis of the strong decays $D^*\to D\pi$ and
electromagnetic decays $D^*\to D\gamma$ \cite{BFO1}, that the
parameter $\lambda$ has the same sign as the parameter
$\lambda'$, which describes the contribution of the
magnetic moment of the heavy (charm) quark. In
the heavy quark limit we have $\lambda'=-1/(6 m_c)$.
Assuming that the finite mass effects are not so large as to
change the sign, we find that $\lambda<0$.
Therefore only four solutions remain. They are shown in
Table 1.

The  calculated  branching 
ratios and polarization variables for the other
semileptonic decays of the type $D\to V$
are in agreement with all the known experimental
data \cite{BFO0} (see Table 2).

In our approach the form factors for $D\to P$ decays are given by 
\cite{BFO0}

\begin{eqnarray}
\label{f1}
{1\over K_{P(i)}} F^{(i)}_{1}(q^2)&=&{1\over f_{P(i)}}
\bigl(-{f_H\over 2}+g f_{H'^*(i)} 
{m_{H'^*(i)} \sqrt{m_H m_{H'^*(i)}}\over 
q^2-m_{H'^*(i)}^2}~\bigr)\;,\\ 
\label{f0}
{1\over K_{P(i)}} F^{(i)}_{0}(q^2)&=&{1\over f_{P(i)}}
\biggl(-{f_H\over 2}-g f_{H'^*(i)}
\sqrt{m_H\over m_{H'^*(i)}}\nonumber\\
&+&{q^2\over m_H^2-m_{P(i)}^2}\bigl[{-f_H\over 2}+
g f_{H'^*(i)} \sqrt{m_H\over m_{H'^*(i)}}\bigr]\biggr)\;.
\end{eqnarray}

\noindent
where the pole mesons and the constants $K_{P(i)}$, which 
contribute to the corresponding processes $D \to PV$ and 
$D \to P_{(1)}P_{(2)}$ are given in in \cite{BFO0}. 
We neglected the lepton mass, so the form factor $F_0$, which 
multiplies $q^\mu$, did not contribute to the decay width.

Using the best known experimental branching ratio -
${\cal B}[D^0\to K^- l^+\nu_l]=(3.68\pm 0.21)\%$
\cite{PDG}, we found two solutions for $g$:

\begin{eqnarray}
\label{solg}
\hbox{SOL.  1 }&:& g \equiv g_{>}=0.15\pm  0.08 \;,\nonumber \\
\mbox{SOL.  2 }&:& g \equiv g_{<}=-0.96\pm   0.18\;.
\end{eqnarray}

\noindent
The quoted error for $g_>$  is mainly due to the uncertainty 
in the value $f_D$, while the quoted error for $g_<$ is mainly 
due to the uncertainty in $f_{D_s^*}$. 
Unfortunately we were not able to choose between the two
possible solutions for $g$ in (\ref{solg}). the branching 
ratios for $D \to P$ transitions are presented in Table 3. 

\vskip 1cm

\centerline{\bf IV. NONLEPTONIC DECAYS}
\vskip 1cm

The effective Hamiltonian for charm decays is given by
\begin{equation}
H_w = \frac{G_F}{{\sqrt 2}} V_{ci} V_{uj}^* 
\{ a_1 ({\bar u}\Gamma_\mu q_j) 
({\bar q}_i \Gamma^\mu c) + a_2 ({\bar u}\Gamma_\mu c) 
({\bar q}_i \Gamma^\mu q_j)\}
\label{hweak}
\end{equation}
where $V_{qq^{\prime}}$ is an element of the CKM matrix, 
 $i$ and $j$ stand for $d$ or $s$ quark flavours, 
$\Gamma_\mu=\gamma_\mu (1-\gamma^5)$, and $a_1$ and $a_2$  
 are the Wilson coefficients: 
\begin{equation}
a_1 = 1.26 \pm 0.04 \qquad a_2 = -0.51 \pm 0.05~.
\label{ai}
\end{equation}
These values are taken from \cite{BURAS,WSB,WSB1} and they are 
in agreement with the next-to-leading order calculation \cite{BURAS}.
The factorization approach in two body nonleptonic decays 
means one can write the amplitude in the form  
\begin{eqnarray}
< AB|  {\bar q}_i\Gamma_\mu q_j {\bar q_k}\Gamma^\mu c | D> 
& = &<A| {\bar q}_i\Gamma_\mu q_j | 0> 
<B| {\bar q_k}\Gamma^\mu c | D> \nonumber\\
& + & <B| {\bar q}_i\Gamma_\mu q_j | 0> 
<A| {\bar q_k}\Gamma^\mu c | D>\nonumber\\
&  + &<AB| {\bar q}_i\Gamma_\mu q_j | 0> 
<0| {\bar q_k}\Gamma^\mu c| D> .
\label{fac}
\end{eqnarray}
In our calculations we take into account only the  
first two contributions. The last one is the annihilation 
contribution, which is absent or negligible in the particular 
decay modes we consider. In other decays  
this contribution was found to be rather important 
\cite{WSB,WSB1,BLMPS}. It was pointed out in 
\cite{WSB,WSB1,BLMMPS,BLP} that the 
simple dominance by the lightest scalar or pseudoscalar mesons 
in $<AB|{\bar q}_i\Gamma_\mu q_j   | 0>$ can not explain the 
rather large contribution present in some of the nonleptonic 
decays, which we will not consider. Our model \cite{BFO0}, being 
rather poor in the number of resonances, is applicable to the 
analysis of the spectator amplitudes, but not the 
annihilation contributions.

We will use the following definitions of the light 
meson and the heavy meson couplings 
$< P(p)| j_{\mu} | 0> = - i f_{P} p_{\mu}$, 
$< V(p,\epsilon^*)| j_{\mu} | 0>  =   m_{V} f_{V} \epsilon^*_{\mu}$, 
$< 0| j_{\mu} | D(P)> = -i f_D m_D v_{\mu}$, and 
$< 0| j_{\mu} | D^*(\epsilon,P)>  = i m_{D^*} f_{D^*}\epsilon_{\mu}$. 
Then using (\ref{defhv}) and (\ref{defhp}) 
we can write the amplitude for the nonleptonic decay 
$D \to PV$ processes as

\begin{eqnarray}
M(D (p) \to P V(\epsilon^*) ) & = & 
\frac{G_F}{{\sqrt 2}} ~\epsilon^* \cdot p~2 m_V
[-w_V K_{V}~ f_P ~A_0(m_P^2)  \nonumber\\
&+ &w_P K_{P}~ f_V ~F_1(m_V^2)]
\label{apv}
\end{eqnarray}
The factors $w_V$, $w_P$, $K_{V}$ and $K_{P}$ 
are given in \cite{BFOP}.

The  $D \to P_1P_2$ decay amplitude is 
\begin{eqnarray}
M(D (p) \to P_{(1)}P_{(2)} ) & = & 
\frac{G_F}{{\sqrt 2}} 
[-i w_1~K_{P(1)}~ f_{P(2)}~ (m_H^2-m_{P(1)}^2) 
~F^{(1)}_{0}(m_{P(2)}^2) \nonumber\\
&-&i w_2  K_{P(2)}~ f_{P(1)}~(m_H^2-m_{P(2)}^2)~ 
F^{(2)}_{0}(m_{P(1)}^2)]
\label{app}
\end{eqnarray}
The factors $w_1$, $w_2$, $K_{P(1)}$ and $K_{P(2)}$ 
are presented in \cite{BFOP}. 

Finally, we find the $D \to V_{(1)}V_{(2)}$ decay amplitude 
to be
\begin{eqnarray}
&&M(D(p) \to  V_{(1)}(p_1,\epsilon_1),V_{(2)}(p_2,\epsilon_2) ) =\hfil \\ 
&&\frac{G_F } {{\sqrt 2}}  
\biggl(w_1 K_{V(1)}~ f_{V(2)}~ m_{V(2)}~ \epsilon_{2\mu}\biggl[
-{2V^{(1)}(m_{V(2)}^2)\over m_H+m_{V(1)}}\varepsilon^{\mu\nu\alpha\beta}
~\epsilon_{1\nu}^*~p_{\alpha}~p_{1\beta}\nonumber\\
&+ &i (m_H+m_{V(1)})~A^{(1)}_{1}(m_{V(2)}^2)~\epsilon_1^{\mu 
*}-i{A^{(1)}_{2}(m_{V(2)}^2) \over m_H+m_{V(1)}}~\epsilon_1^* \cdot p_{V2}~ 
(p+p_{V1})^{\mu}\biggr] \nonumber\\ &+ & w_2 K_{V(2)} ~f_{V(1)}~ m_{V(1)}~ 
\epsilon_{1\mu}\biggl[
-{2V^{(2)}(m_{V(1)}^2)\over m_H+m_{V(2)}}\varepsilon^{\mu\nu\alpha\beta}
~\epsilon_{2\nu}^*~p_{\alpha}~p_{2\beta} \nonumber\\
&+ & i (m_H+m_{V(2)})~A^{(2)}_{1}(m_{V(1)}^2)~\epsilon_2^{\mu *}-  
i{A^{(2)}_{2}(m_{V(1)}^2) \over m_H+m_{V(2)}}~\epsilon_2^* \cdot p_{V1}~ 
(p+p_{V2})^{\mu}\biggr] \biggr) \nonumber 
\label{avv}
\end{eqnarray}
The factors $w_1$, $w_2$, $K_{V(1)}$ and $K_{V(2)}$ for $D\to 
V_{(1)}V_{(2)}$ are given in \cite{BFOP}. 

In order to avoid the strong interaction final state effects in the 
interference between different final isospin states 
we analyze decays in which the final state involves only a single isospin. 
This  occurs when there is an isospin zero particle in the final state 
($\omega$, 
$\Phi$, $\eta$, $\eta '$),  
or when a final state has the maximal third component of the 
isospin; for example, $D^+ \to \bar K^{*0} \pi^+$, 
$D^+ \to \rho^+ \bar K^{*0}$, $D^+ \to \bar K^0 \pi^+$ and 
$D^+ \to \bar K^{*0} \rho^+$ with $\vert I,I_3>=\vert 3/2,3/2>$). 

Our analysis of semileptonic decays $D \to V (P) l \nu_l$ \cite{BFO0} 
left some ambiguity in the choice of the 
model parameters: 
there are two values of $g$, $(g_{<}\enspace {\rm and}\enspace g_{>})$ 
(\ref{solg}) and four solutions for the parameters ($\lambda$, 
$\alpha_1$, $\alpha_2$) (Table 1). 
The calculated nonleptonic decay amplitudes depend on the choice of 
these parameters. However, although the uncertainties 
of the predictions 
are quite large, they are mostly 
due to the calculated errors in  $g_<$ and $g_>$ (\ref{solg}), 
which is in turn due to the uncertainty 
in $f_D$ and $f_{D_s^*}$. 
The only parameter that is not constrained by the semileptonic decay 
data is the parameter $\beta$ in the form factor $A_0$, but 
the predictions for the nonleptonic decay rates are not very sensitive 
to $\beta$. From (\ref{apv}) and (\ref{a0}) it can easily be seen that 
$\beta$ appears multiplied by $m_P^2$ in the $ D \to PV$ decay width 
and is only significant for  the decays $D \to PV$, 
where $P$ is $K$, $\eta$ or $\eta'$.
 
First we discuss the results 
for the decay amplitudes which depend only on the form factors $F_0$ and 
$F_1$ and consequently only on the parameter $g$; namely, 
$D^+ \to \bar K^0 \pi^+$, 
$D^+\to \Phi \pi^+$, $D_s^+\to \rho^+\eta (\eta ')$, 
$D^0\to \Phi\eta $ and $D^0\to \Phi\pi^0$. 
The  comparison with the 
experimental data does not exclude either of the values for $g$, 
$g_<$ or $g_>$ \cite{BFOP}. 
Next, we summarize the results obtained for 
the decays which  depend only on the form factors $V$, 
$A_0$, $A_1$ and $A_2$, and consequently only on the parameters 
($\lambda$, $\alpha_1$, 
$\alpha_2$); namely, $D_s^+ \to \Phi \pi^+$, $D_{s}^+\to \Phi \rho^+$,  
$D^0\to \Phi \rho^0$ and $D^+ \to \bar K^{*0} \rho ^+$. 
The decay $D_s^+ \to \Phi \pi^+$ depends also 
on the parameter $\beta$, but this dependence is very slight, 
since the light pseudoscalar meson in the final state is a $\pi$. 

The results for all sets are in rather 
good agreement with the experimental data, with the exception of 
$D^0\to\Phi\rho^0$, which we do not understand.  

In addition to the above two types of nonleptonic decays, 
there are two measured branching ratios for $D^+\to\bar K^{*0}\pi^+$ 
and $D^+\to \rho^+ \bar K^0$. Their decay amplitudes depend on 
both $g$ and the parameters $\lambda$, $\alpha_1$, $\alpha_2$. 
The branching ratio for $D^+\to\bar K^{*0}\pi^+$, which is not sensitive 
to $\beta$ since the $\pi$ mass is small, excludes the parameter $g_<$, 
the sets II and IV, and  prefers 
$g=g_>=0.15\pm 0.08$ and the set I. 

>From the $D^+\to \rho^+ \bar K^0$ decay, which has $K$ pseudoscalar meson 
in the final state, one can then  estimate the  parameter $\beta$. 
Unfortunately, this decay has a considerable 
experimental error, $BR=(6.6 \pm 2.5) \%$ \cite{PDG}, which results in  
large error in $\beta$:

\begin{equation}
\beta=3.5\pm 3~\;.
\label{bet}
\end{equation}

The predictions for the branching ratios for the possible 
decays are presented in Table 4  assuming set I for $\lambda$, 
$\alpha_1$ and $\alpha_2$, $g=g_>=0.15\pm 0.08$ and $\beta=3.5\pm 3$.  
The quoted errors are due 
to the uncertainties in the model parameters, mainly $g$.

\vskip 1cm
 
\centerline{\bf VI. SUMMARY}

\vskip 1cm               
We have proposed a method to include the light vector meson resonances in 
the weak currents using HQET and CHPT. Instead of the propagators used in 
HQET we have used full propagators for the intermediate 
heavy meson states. In this way we obtain a pole-type behavior 
of the form factors for the matrix elements of the vector currents, and 
a constant behavior of the form factors of the axial current.
The calculated branching ratios are in agreement with the 
experimental results. We have predicted  the other semileptonic 
decays that have not yet been observed.
In addition we have calculated the branching ratios for  the nonleptonic 
decay modes 
$D \to P V$, $D \to P_1 P_2$ and $D \to V_1 V_2$  
in which the annihilation contribution is absent or negligible, 
and  in which the final state involves only a single isospin in 
order to avoid the effects of strong interaction phases. 
Factorization of the 
matrix elements was then assumed and we used the effective 
model developed to describe the semileptonic decays $D\to V (P) l \nu_l$ to 
calculate the nonleptonic matrix elements. 
We reproduced the experimental results for branching 
ratios for the $D^+ \to \bar 
K^{*0} \pi^+$, $D^+ \to \rho^+ \bar K^{0}$, $D_s^+\to \Phi\pi^+$, $D_s^+\to 
\rho^+\eta$, $D^+ \to \bar K^0 \pi^+$, $D_s^+\to \Phi\rho^+$ and 
$D^+\to \bar 
K^{*0}\rho^+$ decays, albeit within substantial uncertainties. 
We also determined the set of parameters $\lambda$, 
$\alpha_1$ , $\alpha_2$ and $g$, 
which gave the best agreement with the experimental results and  
used this set of parameters to estimate the parameter $\beta$ 
from the branching ratio for $D^+ \to \rho^+ {\bar K}^{0}$.  
We then made the predictions for a number of  
nonleptonic decay rates which have not yet been measured.

\begin{table}[ht]
\begin{center}
\begin{tabular}{|c|c|c|c|}\hline
& $\lambda$ [GeV$^{-1}$]
& $\alpha_1$ [GeV$^{1/2}$]
& $\alpha_2$ [GeV$^{1/2}$]  \\ \hline
Set 1 & $-0.34 \pm 0.07$ & $-0.14 \pm 0.01$ &
$-0.83 \pm 0.04$\\
Set 2 & $-0.34 \pm 0.07$ & $-0.14 \pm 0.01$ &
$-0.10 \pm 0.03$\\
Set 3 & $-0.74 \pm 0.14$ & $-0.064 \pm 0.007$ &
$-0.60 \pm 0.03$\\
Set 4 & $-0.74 \pm 0.14$ & $-0.064 \pm 0.007$ &
$+0.18 \pm 0.03$\\ \hline
\end{tabular}
\label{tabset}
\caption{Four possible solutions for the model parameters
as determined by the $D^+\to\bar{K}^{*0}l^+\nu_l$ data.}
\end{center}
\end{table}

\begin{table}[ht]
\begin{center}
\begin{tabular}{|c||c|c|c|}
\hline
decay & ${\cal B}$ [$\%$] & $\Gamma_L/\Gamma_T$ & $\Gamma_+/\Gamma_-$ \\
\hline
\hline
$D^0\to K^{*-}$ & $1.8 \pm 0.2$ & $1.23 \pm 0.13$ & $0.16 \pm 0.04$ \\
                & $(2.0 \pm 0.4)$&                 &                 \\
\hline
$D_s^+\to\Phi$ & $1.7 \pm 0.1$ & $1.2 \pm 0.1$ & $0.16 \pm 0.04$ \\
               & $(1.88\pm 0.29)$&$(0.6 \pm 0.2)$ &                 \\
\hline
$D^0\to\rho^-$ & $0.17\pm 0.02$ & $1.34\pm 0.2$ & $0.15\pm 0.10$ \\
\hline
$D^+\to\rho^0$ & $0.22\pm 0.02$ & $1.4\pm 0.2$ & $0.15\pm 0.10$ \\
               & $(<0.37)$      &                               \\
\hline
$D^+\to\omega$ & $0.21\pm 0.02$ & $1.4\pm 0.2$ & $0.16\pm 0.10$ \\
\hline
$D_s^+\to K^{*0}$ & $0.17\pm 0.02$ & $1.3\pm 0.2$ & $0.15\pm 0.10$ \\
\hline
\end{tabular}
\end{center}
\label{tabfour}
\caption{The branching ratios and polarization ratios for the
$D\to V$ semileptonic decays. Where available,
the experimental data is quoted in brackets.}
\end{table}

\begin{table}[ht]
\begin{center}
\begin{tabular}{|c||c|c|c|}
\hline
decay & ${\cal B}_1$ & ${\cal B}_2$ & exp. \\
\hline
\hline
$D^+\to{\bar K}^0$ & $9.4 \pm 0.5$ & $9.4 \pm 0.5$ & $6.7 \pm 0.8$ \\
\hline
$D_s^+\to\eta$ & $3 \pm 3$ & $2 \pm 2$ & \\
\hline
$D_s^+\to\eta'$ & $1.6 \pm 0.7$ & $0.9 \pm 0.5$ & \\
\hline
$D_s^+\to(\eta+\eta')$ & $4 \pm 3$ & $3 \pm 3$ & $7.4 \pm 3.2$ \\
\hline
$D^0\to\pi^-$ & $0.47 \pm 0.05$ & $0.5 \pm 0.5$ & $0.39^{+0.23}_{-0.12}$ \\
\hline
$D^+\to\pi^0$ & $ 0.59 \pm 0.06$ & $0.7 \pm 0.6$ & $0.57 \pm 0.22$ \\
\hline
$D^+\to\eta$ & $0.18 \pm 0.05$ & $0.1 \pm 0.2$ & \\
\hline
$D^+\to\eta'$ & $0.021 \pm 0.005$ & $0.01 \pm 0.01$ & \\
\hline
$D_s^+\to K^0$ & $0.4 \pm 0.2$ & $0.2 \pm 0.3$ & \\
\hline
\end{tabular}
\end{center}
\label{tabsix}
\caption{The branching ratios for the $D\to P$ semileptonic decays.}
\end{table}

\begin{table}[ht]
\begin{center}
\begin{tabular}{|c||c|c|c|c||c|}
\hline
decay & ${\cal B}_{th}[\%]$  & ${\cal B}_{exp}[\%]$ \\
\hline
\hline
$D^+ \to \bar K^{*0} \pi^+$ & $2.4\pm 1.2$ & $1.92\pm 0.19$ \\\hline
$D^+ \to \rho^+\bar K^{0} $ & $6.6\pm 3.0$ & $6.6\pm 2.5$ \\\hline
$D^+ \to \Phi \pi^+$ & $0.40\pm 0.12$ & $0.61\pm 0.06$\\\hline
$D_s^+ \to \Phi \pi^+$ & $5.4\pm 0.5$ & $3.6\pm 0.9$\\\hline
$D_s^+ \to \rho^+\eta$  & $9.0\pm 2.5$ & $10.3\pm 3.2$\\\hline
$D_s^+ \to \rho^+\eta '$ & $4.5\pm 1.3$ & $12.0\pm 4.5$\\\hline
$D^+ \to \bar K^0 \pi^+$ &  $2.2\pm 0.7$ & $2.74\pm 0.29$\\\hline
$D_s^+ \to \Phi \rho^+ $ & $4.4\pm 0.8$  & $6.7\pm 2.3$\\\hline
$D^0 \to \Phi \rho^0$ & $0.029\pm 0.005$ &  $0.11\pm 0.03$\\\hline
$D^+ \to \bar K^{*0}\rho^+$ & $2.9\pm 0.4$ & $2.1\pm 1.4$\\\hline
$D^+ \to \rho^+\eta$  & $0.05\pm {0.9 \atop 0.05}$ & $<1.2$\\\hline
$D^+ \to \rho^+\eta '$ & $0.02\pm {0.2\atop 0.02}$ & $<1.5$\\\hline
$D^0 \to \Phi\eta$  & $0.018\pm 0.005$ & $<0.28$\\\hline
$D^0 \to \omega\eta$  & $0.09\pm 0.03$ & $-$\\\hline
$D^0 \to \omega\eta '$ & $0.015\pm 0.015$ & $-$\\\hline
$D^0 \to \Phi\pi^0$ & $0.07\pm 0.02$ & $<0.14$\\\hline
$D^+ \to \Phi \rho^+$ & $0.14\pm 0.03$ &  $<1.5$\\\hline
$D^0 \to \Phi \omega $ & $0.028\pm 0.004$ &  $<0.21$\\
\hline 
\end{tabular} 
\end{center}
\label{res}
\caption{The predicted (column two) and measured (column three) 
branching ratios for the nonleptonic decay modes. }
\end{table}
\newpage

\end{document}